\documentclass[reprint, aip, cha]{revtex4-1}

\usepackage{bm,graphicx,color,amssymb,amsmath,amsthm,lineno}
\usepackage{hyperref}%

\begin{document}

\title{FliI$_{6}$-FliJ molecular motor assists with unfolding in the type III secretion export apparatus}
\author{Jiri Kucera and Eugene M. Terentjev}
 \email{emt1000@cam.ac.uk}
\affiliation{Cavendish Laboratory, University of Cambridge, Cambridge, CB3 0HE, U.K.}

\date{\today}

\begin{abstract}
\noindent \textbf{Abstract}. \ The role of rotational molecular motors of the ATP synthase class is integral to the metabolism of cells. Yet the function of FliI${}_6$-FliJ complex - a homolog of the $F_1$ ATPase motor - within the flagellar export apparatus remains unclear. We use a simple two-state model adapted from studies of linear molecular motors to identify key features of this motor. The two states are the 'locked' ground state where the FliJ coiled coil filament experiences fluctuations in an asymmetric torsional potential, and a 'free' excited state in which FliJ undergoes rotational diffusion. Michaelis-Menten kinetics was used to treat transitions between these two states, and obtain the average angular velocity of the FliJ filament within the FliI$_{6}$ stator: $\omega_{max}\approx 9.0\:$rps. The motor was then studied under external counter torque conditions in order to ascertain its maximal power output: $P_{max}\approx 42\:$k$_B$T/s, and the stall torque: $G_\mathrm{stall}\approx 3\:$k$_B$T/rad. Two modes of action within the flagellar export apparatus are proposed, in which the motor performs useful work either by continuously 'grinding' through the resistive environment, or by exerting equal and opposite stall force on it. In both cases, the resistance is provided by flagellin subunits entering the flagellar export channel prior to their unfolding. We therefore propose that the function of the FliI${}_6$-FliJ complex is to lower the energy barrier and therefore assist in unfolding of the flagellar proteins before feeding them into the transport channel.
\end{abstract}

\maketitle

\section*{Introduction}

With the advance of imaging techniques, our view of living matter and of its fundamental units, the cells, has changed dramatically. These micron-sized `bags of chemicals' turned out to be run by complex yet physically describable networks of proteins, lipids, and carbohydrates. The immense number of processes occurring in a cell at any given moment would be unattainable in such a packed environment without a considerable level of organization and reaction catalysis. In general, this is achieved by proteins -- long chains of amino acids that come in various sizes and shapes when folded in solution. Their functionality originates from polarity and hydrophobicity of different aminoacids, and is responsible for the unique self-assembly and the resulting properties that range from simple structural support to powerful catalysers. 	

One such class of proteins are the molecular motors \cite{stryer_1999, kolomeisky2007molecular, julicher1997modeling}. These large molecular complexes are responsible for organised powered movement within cells and can be characterised by the following properties. They consume energy (usually chemical energy stored in molecules of ATP, or in electrochemical gradient of ions across membranes) and transform it into mechanical work. The energy input is crucial to drive the system out of equilibrium.  Another requirement to achieve directed motion is the presence of asymmetry (or broken symmetry) in the underlying potential governing the motion of the motor. Lastly, the microscopic nature of the motors and the surrounding heat bath means that their motion is inherently stochastic and is therefore subject to overdamped thermal fluctuations. 

Several types of motors are distinguished. Cytoskeletal motors move along polarised tracks defined by rigid filaments: myosin along actin fibres, while kinesin and dynein along microtubules. Polymerisation motors, on the other hand, output mechanical work by elongating themselves. In doing so they might, for instance, change the shape of the cell \cite{alberts_2015}. Lastly, rotary motors convert chemical energy into rotational motion and are the main subject of this study.

One of the most important and most studied rotary motors is the $F_1 F_0$ ATP synthase complex (ATPase) \cite{abrahams1994structure,yoshida2001,junge2009torque}. It is difficult to overstress its importance in almost any living organism. The ATPase is embedded in the mitochondrial membrane. When sugars are burned in mitochondria, an $H^+$ ion gradient is set up across the membrane. The  $F_0$ motor of ATPase then harnesses the electrochemical energy stored in this gradient to rotate one of its components \cite{oster1998f0}. In doing so, ADP (adenosine diphosphate) is converted into highly energetic ATP (adenosine triphosphate) at the expense of the $H^+$ gradient dissipation. ATP is subsequently used as the main fuel for thousands of reactions in the cell releasing ADP in the process. The other function of the ATPase is to recycle the ADP by phosphorylating it back into ATP by the $F_1$ motor of the complex.

The structure of the F-ATPase is very well known \cite{abrahams1994structure, gibbons2000structure}. It consists of two main domains. The $F_0$ domain is embedded in the membrane, and is responsible for converting the energy stored in the ion gradient into rotation. The $F_1$ domain sits at the other end of the same drive-shaft (called $\gamma$-subunit) and works as a stator, converting rotational energy provided by the $F_0$ into the binding energy of ATP. The $\gamma$-shaft  transfers the torsional mechanical energy from one domain to the other. 

However, each domain can function on its own. Thus, the $F_1$ motor can instead consume ATP to rotate the $\gamma$-shaft against the $F_0$, and force the $H^+$ ions across the membrane to build up their concentration gradient. Independent action of the $F_0$ motor has been extensively observed in-vitro \cite{noji1997,junge2008,noji2011}, and much useful information about the underlying molecular processes was obtained in this way. Although the hexamer structure of the $F_0$ (and its many homologues, such as $V_0$-ATPase or FliI${}_6$ complex) is very well known, the physical mechanism of the $F_1$ motor action is less clear.  Some of the prevailing theories are based on the idea of a ``power stroke" \cite{spetzler2009single, martin2014anatomy}. In this model, an asymmetry in the $\alpha$ and $\beta$ units of the $F_0$ hexamer complex pushes, upon conformation change, on the base of the $\gamma$-shaft and thus exerts a torque. In fact, the majority of publications \cite{sakaki2005one, hornung2008determination} accept the notion of a constant torque being exerted by the motor, acting against friction, to justify the observed rotational velocity. However, this picture has several faults. Firstly, the scale of the system and the highly viscous environment surrounding the complex imply that it operates in the overdamped regime. This means that any inertial effects must be neglected. As a result, the picture of a ``turbine" pushed by a stream of protons, or a shaft spun by a ``cog" has to be replaced with a fully stochastic construction subject to Brownian motion. Secondly, high resolution measurements of the rotational motion clearly show the possibility of a reverse step \cite{yasuda1998f1}:  a phenomenon that cannot be accounted for in the power stroke model. Lastly, numerous studies \cite{yasuda1998f1, junge2009torque} use the false concept of torque generation to arrive at efficiencies close to 100\%. This is, of course, at odds with classical thermodynamics and reversible Carnot engines. This all illustrates the lack of understanding when it comes to the $F_0$-ATPase dynamics. Frasch et al. proposed that the rotation is partially derived from elastic energy stored in the $\gamma$-shaft \cite{martin2018elastic}. This idea was further developed by Kulish et al. into a stochastic two-level model \cite{kulish2016f}. Here we adopt this methodology and apply it onto a different molecular motor in the hope of predicting some of its properties. 

\begin{figure}
	\centering
	\includegraphics[width=0.36\textwidth]{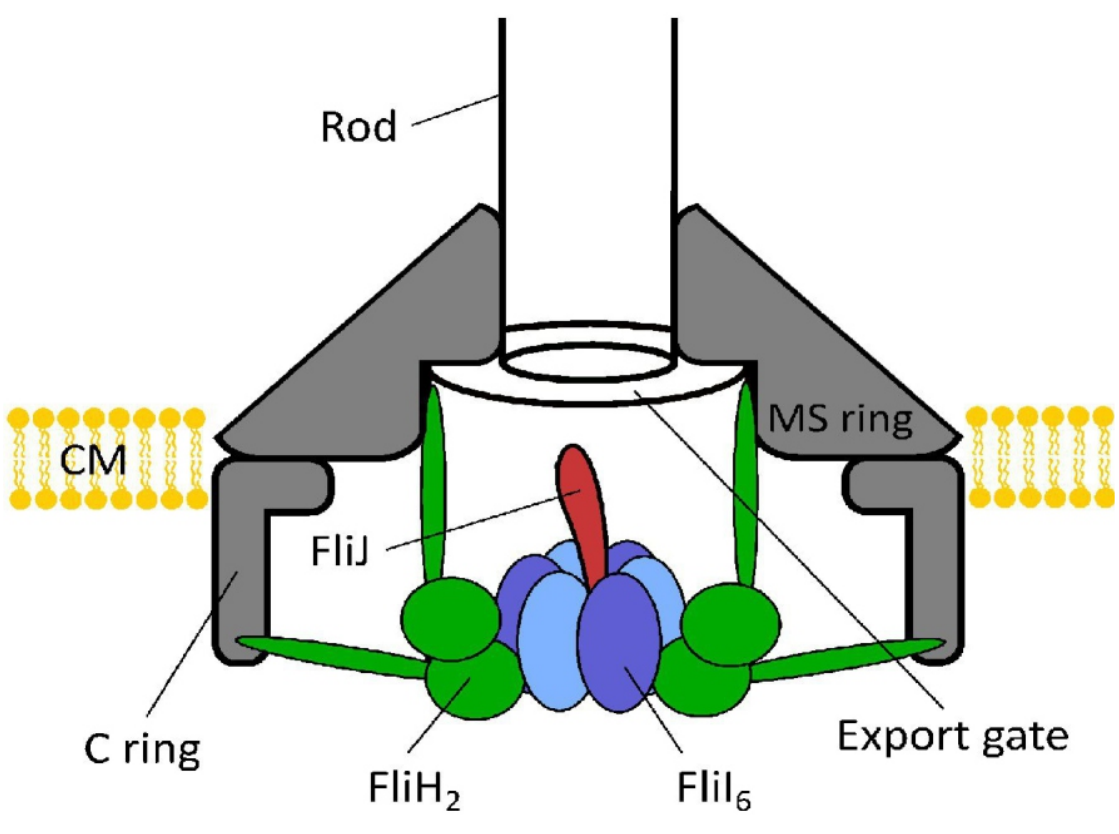}
	\caption{Schematic picture of the flagellar export apparatus \cite{minamino2014bacterial}, with the FliI${}_6$-FliJ complex at the entrance to the export channel.  The sketch is based on the 3D reconstruction of the apparatus from crystallographic data \cite{abrusci2013architecture}. It has been shown \cite{ibuki2013interaction} that FliJ makes direct contact with the export gate, suggesting a coupling between the ATPase activity of FliI and substrate funnelling into the export channel.}
	\label{exportaparatus}
\end{figure}

Structural features of the $F_1$ ATPase are typical of a whole class of molecular motors that are generically called the `ATPases'. They all contain a hexamer stator which imparts the rotation on a coiled-coil filament. One such ATPase, which is arguably a much earlier evolutionary construction than the `modern' F-ATP synthase, is involved in the type III secretion export apparatus that facilitates assembly of bacterial flagella. It has been shown that such an ATP-driven rotary motor is an integral part of the flagellar export mechanism \cite{abrusci2013architecture, minamino2014bacterial, bai2014assembly}. This motor is called the FliI${}_6$-FliJ complex, and it sits at the bottom of the export channel (fig. \ref{exportaparatus}). FliI${}_6$ is a hexamer of identical protein subunits FliI that corresponds closely to the  $\alpha_3 \beta_3$ stator complex in $F_1$ motor (PDB: 2JDI), whereas FliJ is a coiled-coil filament that resembles the $\gamma$-shaft (PDB: 1D8S)  \cite{hausrath1999}. Both were shown to be very closely evolutionarily related \cite{imada2007structural, ibuki2011common, kishikawa2013common}. However, the exact purpose of this complex in the export apparatus is not firmly established. 

Here we develop a physical model of ATP-induced rotation of this motor, and use it to ascertain the function of the FliI${}_6$-FliJ complex in the flagellar export -- which we believe to be in the `assisted unfolding' of protein subunits (FliC and FlgD-E) that are later fed into the flagellum channel to be transported to the growing distal end \cite{evans2013chain, renault2017bacterial}. We present a simplified two-state continuum model in which the motor moves by rotationally diffusing in two distinct potential landscapes, one of which originates from elastic properties of the system with an underlying assymetry that gives rise to directed motion. Transition between the two states is induced by the ATP hydrolysis which we describe using Michaelis-Menten reaction kinetics. First, we look closely at the structure of the motor complex in order to ascertain the form of the two potentials.

\section{$\mathrm{\bf{FliJ}}$ - the ATP$\mathrm{\bf{ase}}$ coiled coil}

\begin{figure}
	\centering
	\includegraphics[width=0.3\textwidth]{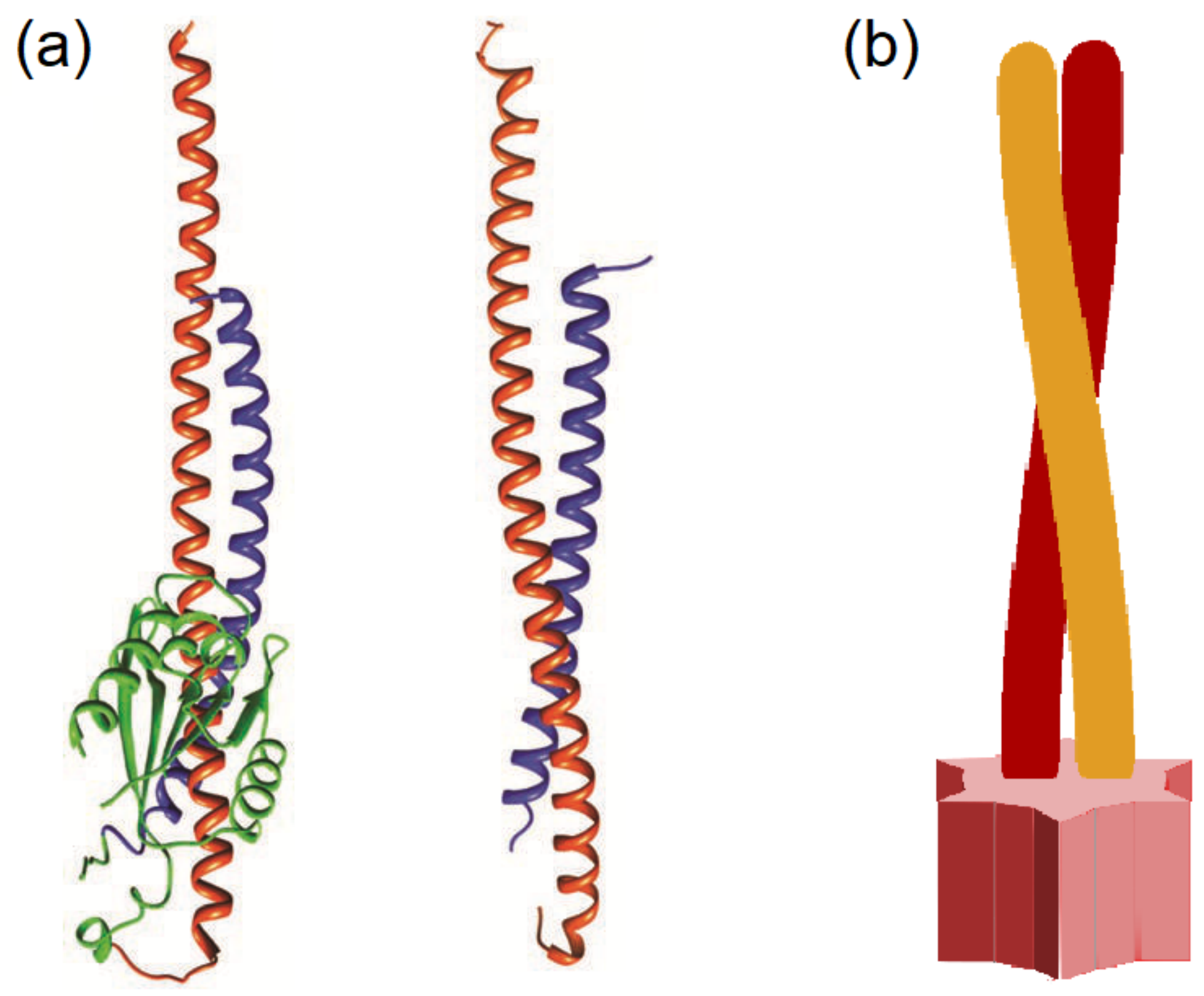}
	\caption{The FliJ shaft. (a) Comparison of the $\gamma$-shaft  in the F-ATPase (left, PDB: 1D8S) with the structure of FliJ (right, PDB: 3AJW). Notice the similarity between the coiled coil patterns found in both proteins \cite{ibuki2011common}. (b) Schematic picture of the FliJ coiled coil. The structure at the base of the filament fits into the FliI stator in a lock-key fashion, and prevents the filament from rotating in one of the motor states.}
	\label{coiledcoil}
\end{figure}

As mentioned above, FliJ is homologous to the $\gamma$-shaft of the F-ATPase, see fig. \ref{coiledcoil}. It binds to the central cavity of the FliI${}_6$ hexamer stator and performs the rotational motion. The interaction of the ``tooth" protrusions (fig. \ref{coiledcoil}b) with matching grooves in the stator cavity ensures that the shaft can be in two states: one fixed, the other free to rotate. The unit itself is composed of two $\alpha$-helices intertwined together to form a coiled coil. We shall now discuss the elastic properties of this filament that govern the motion of the motor.

\begin{figure}
	\centering
	\includegraphics[width=0.48\textwidth]{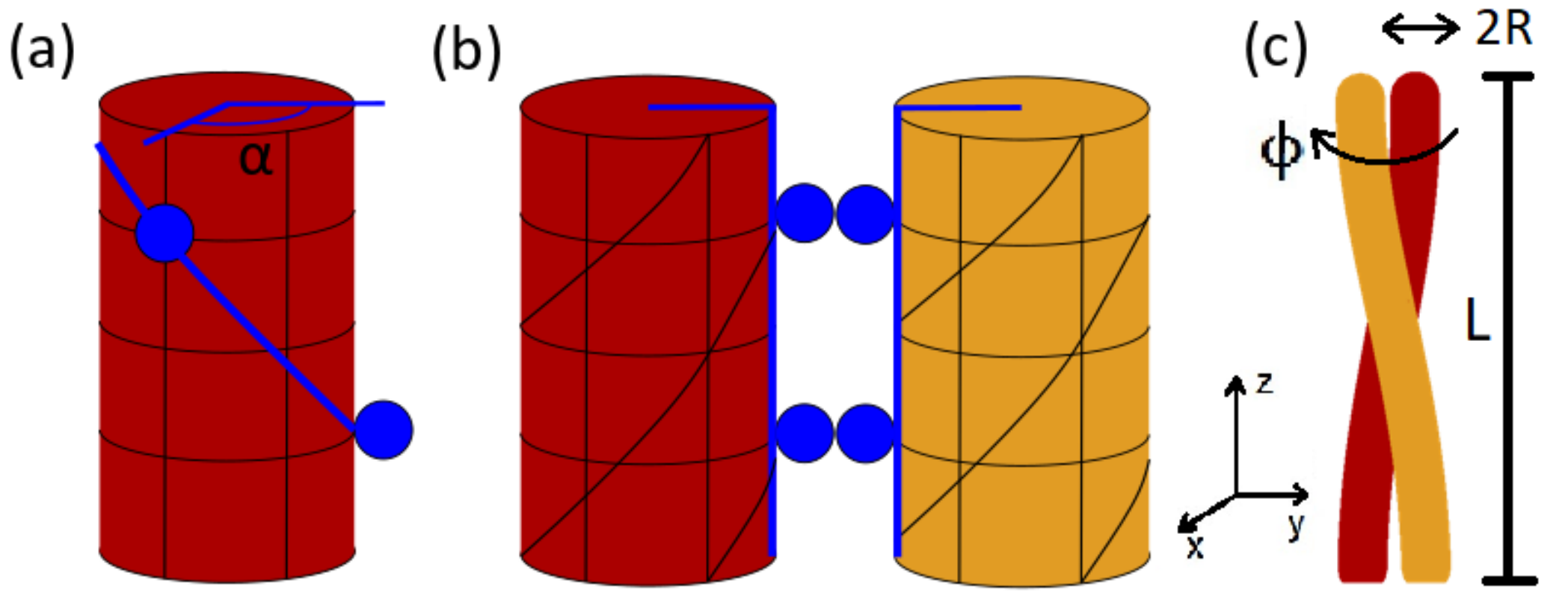}
	\caption{Coiled coil geometry \cite{neukirch2008chirality}. (a) Simplified $\alpha$-helix. Hydrophobic residues are periodically spaced along the helix and are depicted in dark blue. For the heptad repeat \cite{cohen1990,wolgemuth2008elasticity}, the polar angle between subsequent residues is $\alpha\approx20^\circ$. Blue line connecting hydrophobic residues will be referred to as the ``seam line''. (b) When the hydrophobic residues lock to each other, the seam line straightens out, twisting the $\alpha$-helices in the process. (c) The helices wrap around each other to reduce their twisting energy whilst maintaining a straight seam line. $R\approx 4.8\AA$ is the radius of a single helix as well as of the coiled coil structure (assuming tight contact of the helices). $L\approx78\AA$ is the length of the coiled coil region in FliJ, and the angle $\phi$ measures the overall twisting of the coiled coil away from its equilibrium. All numerical values are taken from the Protein Data Bank.}
	\label{elasticity}
\end{figure}

\begin{figure}
	\centering
	\includegraphics[width=0.33\textwidth]{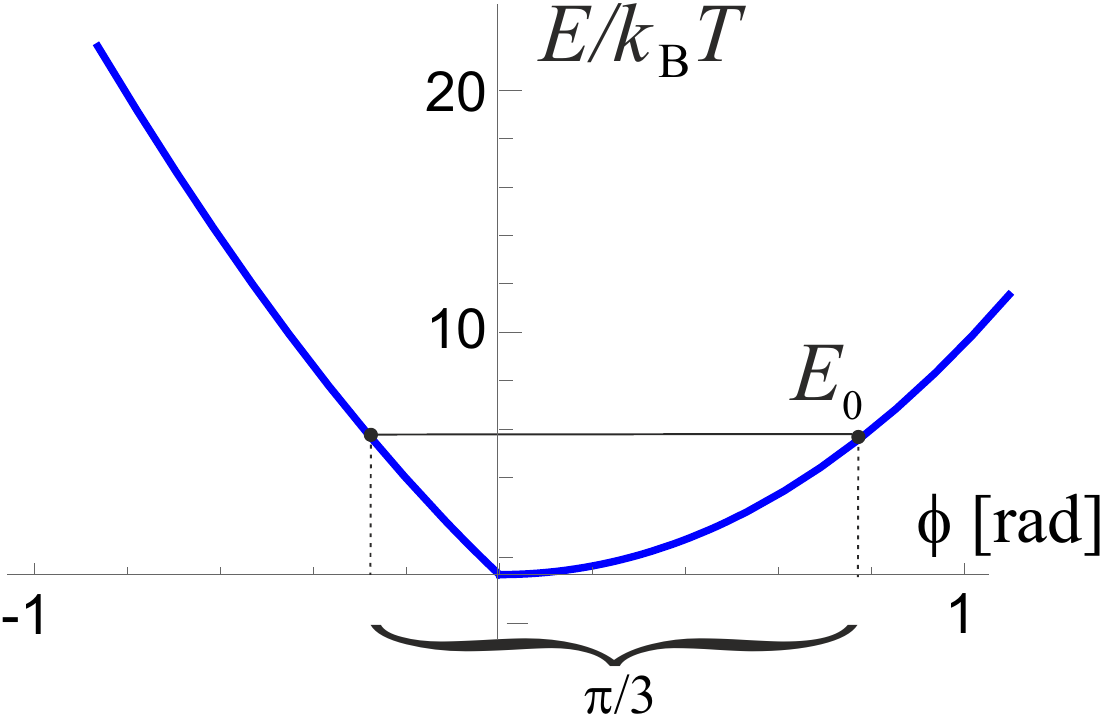}
	\caption{The overall energy of the coiled coil as a function of the twisting angle. Notice the resulting asymmetry which is a key feature of all molecular motors. The period of the hexamer is $\Delta \phi = \pi/3$, which determines the asymmetry parameters of the potential, $a$ and $b$, and the energy barrier $E_0$ (cf. fig. \ref{motoraction}).}
	\label{egraph}
\end{figure}

$\alpha$-helix is a protein folding motif that consists of a long chain of amino acids that spontaneously assembles into a right-handed helix. The persistence length of a typical alpha helix is about 100nm \cite{cohen1990,choe2005elasticity, palenvcar2014buckling, wolgemuth2008elasticity}. For two helices to form a coiled coil  there needs to be a favourable interaction that balances out the energy needed to twist and bend the individual $\alpha$-helices. This originates from the hydrophobic interaction of matching side chains that protrude from the helices at regular intervals (fig. \ref{elasticity}a). The hydrophobic binding energy is an order of magnitude larger than the elastic energy \cite{stryer_1999, alberts_2015} and will, therefore, govern the equilibrium conformation of the helices. As the overlap between the sidechains is maximized, the helices wrap around each other (fig. \ref{elasticity}c). The ``seam line'' connecting the repeated matching hydrophobic residues is straightened in the process (fig. \ref{elasticity}b-c). If we treat an $\alpha$-helix as a cylinder, then its elastic energy is given by \cite{landau_lifshitz_2008}:
\begin{equation}
\int \frac{1}{2}EI\left(\left(\frac{d^{2}Y}{dz^{2}}\right)^2 + \left(\frac{d^{2}X}{dz^{2}}\right)^2\right) + \frac{1}{2}C\psi^2 \mathrm{d}z
\label{el}
\end{equation}
The first term corresponds to bendin, and the second term corresponds to twisting of the cylinder. $X$ and $Y$ are the coordinates of the centre of the helix in cartesian coordinates. $E$ denotes the Young modulus, so that $B=EI$ is the bending modulus, while $C$ is the twisting modulus. The last parameter that needs to be addressed is the twisting angle $\psi$ of a single helix. When the residues align (fig. \ref{elasticity}b), the cylinders acquire a pitch $\psi_{0}={\alpha}/{ph}$ where $\alpha$ is the angle between subsequent hydrophobic residues, $h=1.5\AA$ is rise per amino acid, and $p$ is the period of hydrophobic residues along the helix. The period affects $\alpha$ as well as the handedness of the coiled coil. FliJ is a left-handed coiled coil which limits $p$ to a few allowed values  \cite{neukirch2008chirality}. Based on observed equilibrium twisting angles the most suitable value is $p=7$, the so-called heptad repeat, which gives $\alpha=20^\circ$. The cylinders then intertwine in order to reduce the overall pitch by $\phi/L_{0}$ where $\phi$ is the coiled coil twisting angle and $L_{0}$ is contour length of a single $\alpha$-helix. The relation between helix contour length $L_{0}$ and coiled coil length $L$ is given by $L=\sqrt{L_{0}-(R\phi)^2}$. By noting that $B$ and $C$ are both related to the persistence length $l_{p}$ by $B=k_BTl_{p}\approx 2C$ \cite{wolgemuth2008elasticity} we can integrate Eq. \eqref{el} to obtain the expression for elastic energy of the the two-helix coiled-coil system:
\begin{equation}  \label{el2}
E_\mathrm{el}=k_BTl_{p}\left(\frac{L_{0}}{2}\left(\frac{\alpha}{ph}-\frac{\phi}{L_{0}}\right)^{2}+LR^{2}\left(\frac{\phi}{L}\right)^4\right)
\end{equation}
\begin{figure}
	\centering
	\includegraphics[width=0.4\textwidth]{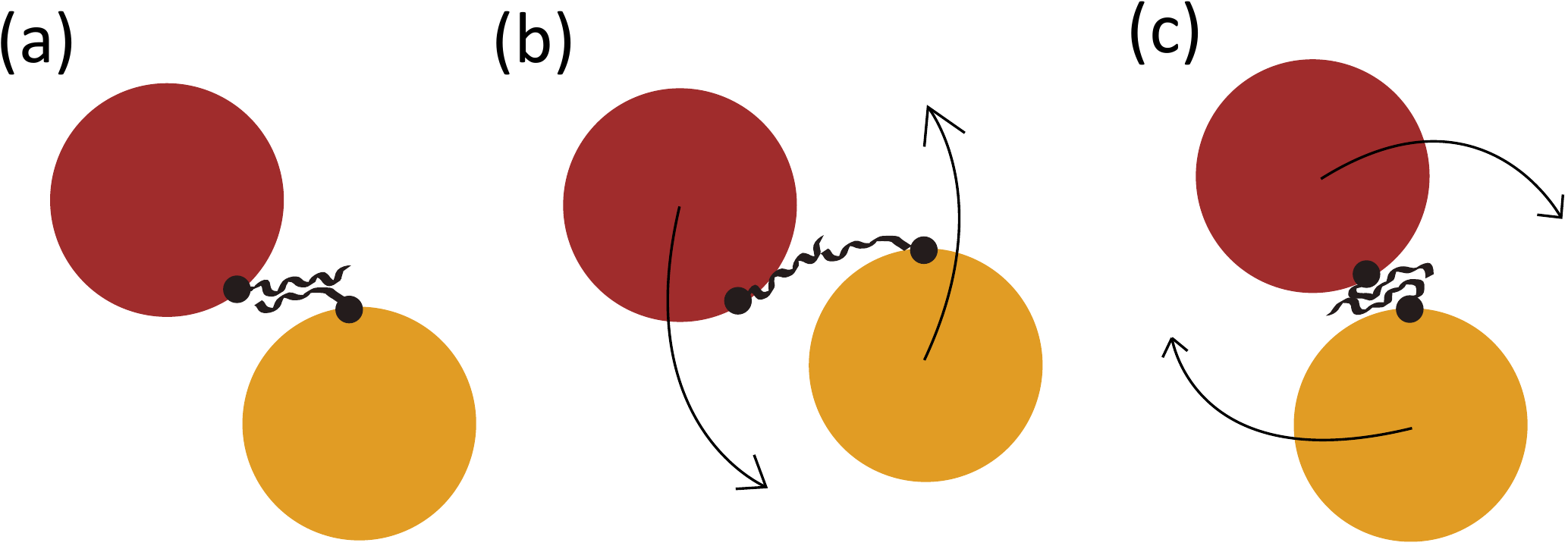}
	\caption{Hydrophobic interaction. (a) In equilibrium ($\phi=0$), the side chains are in tension and fully overlapped. (b) When $\phi<0$, the amino acids are further apart and overall energy rises. (c) When $\phi>0$, the side chain conformation changes but due to their flexibility the extent of overlap remains approximately the same so $E\approx0$.}
	\label{sidechains}
\end{figure}
In order to describe the energy landscape around the equilibrium twisting angle, we also need to consider perturbation to the hydrophobic seam energy. The average energy associated with hydrophobic interaction is roughly $U_{0}=1.7k_BT$ per methylene group \cite{stryer_1999}. The average number of carbon atoms in a side chain is $N\approx5$ with length $d=7.7\AA$. The length of the shorter helix that composes the FliJ unit is $L\approx78\AA$ which means there will be 7 interacting residues along the coiled coil. The overall conformation is a result of competition between elasticity and hydrophobicity, so in equilibrium the interacting hydrophobic residues are in tension, preventing the coiled coil from untwisting (fig. \ref{sidechains}a). When the filament is twisted in one direction the residues are moved further apart, reducing their overlap (fig. \ref{sidechains}b). Twisting in the other direction has no effect for relatively small twisting angles because the side chain residues themselves are flexible (fig. \ref{sidechains}c). The hydrophobic energy as a function of twisting angle is therefore asymmetric and has approximately the following form:
\begin{align}
E_\mathrm{hph}&\approx-\sum_{n=1}^{7} \frac{2\phi R}{d} \frac{nph}{L_{0}}NU_{0} &&\quad \phi<0 \label{hp1} \\
E_\mathrm{hph}&\approx0 &&\quad \phi\ge0  \nonumber
\end{align}
The first fraction $(2\phi R/d)$  in Eq. \eqref{hp1} assumes a linear scaling of energy with the chain overlap, and the second fraction $(nph/L_0)$ takes into account different displacement of chains along the helices. Combining equations \eqref{el2} and \eqref{hp1} we arrive at the final form of the potential energy around equilibrium twisting angle, which is shown in fig. \ref{egraph}. 

\section{$\mathrm{\bf{FliI}}_{6}$ - the ATP$\mathrm{\bf{ase}}$ stator}
\begin{figure}
	\centering
	\includegraphics[width=0.35\textwidth]{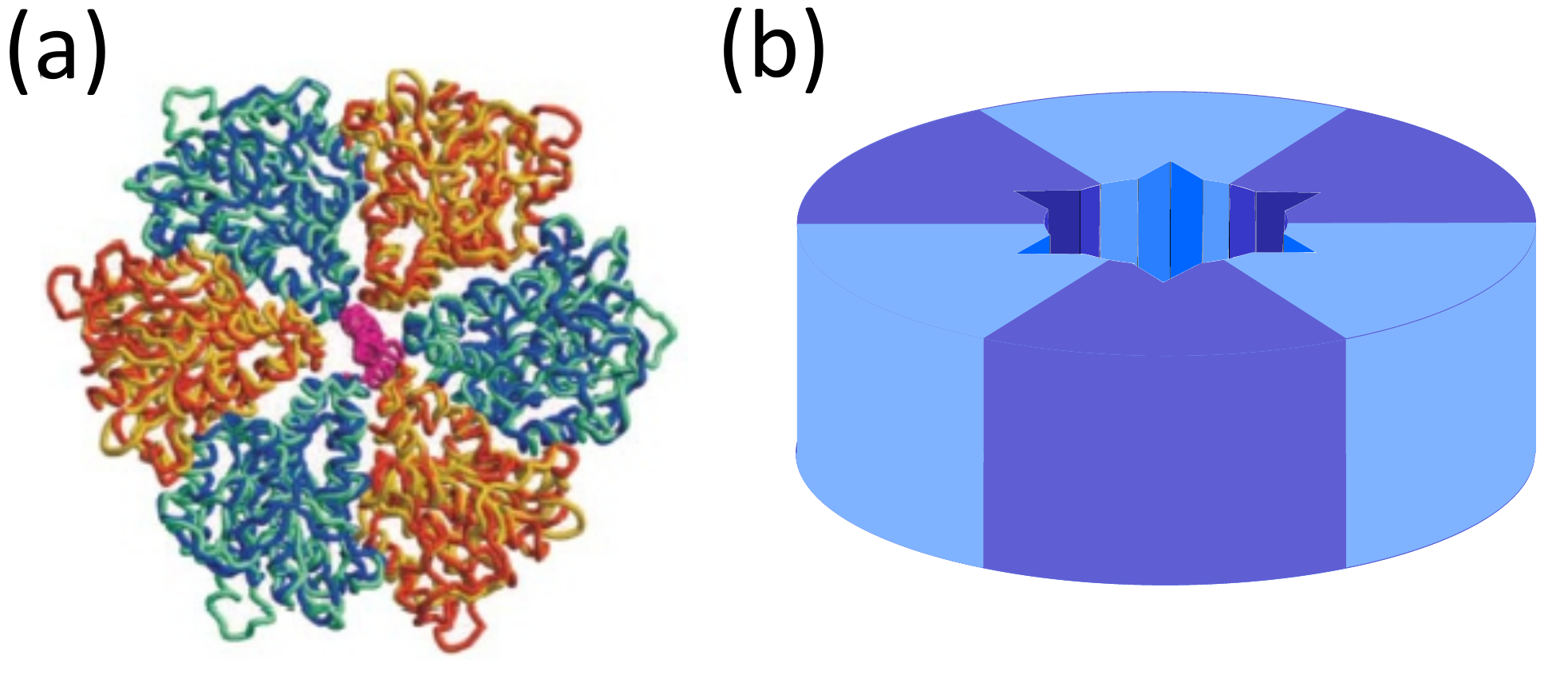}
	\caption{The ATPase stator. (a) A top view of the FliI${}_6$-FliJ complex \cite{imada2007structural}. The equivalent FliI subunits are shown in alternating blue and orange. The central cavity is housing the FliJ shaft shown in purple. (b) Schematic picture of the FliI${}_6$ stator, illustrating the binding sites for the FliJ in the central cavity, represented by triangular grooves. A ``tooth'' protrusion situated on the base of FliJ fits into this groove and locks it in place (see below).}
	\label{hexamer}
\end{figure}

In solution, individual FliI units spontaneously assemble to form a hexamer with a six-fold symmetry and a central cavity \cite{claret2003oligomerization}. This structure forms a stator into which the FliJ coiled-coil ``shaft''  is anchored (fig. \ref{hexamer}a). Each individual subunit is capable of binding to ATP and hydrolyse it to ADP. Overall, the FliI${}_6$ shows very similar ATP activity to the $\alpha_3 \beta_3$ complex of the $F_1$ motor, with the only difference of having 6 binding sites as opposed to 3. In our model, each FliI unit also contains a steric binding site for the FliJ shaft (fig. \ref{hexamer}b). 

\begin{figure}
	\centering
	\includegraphics[width=0.48\textwidth]{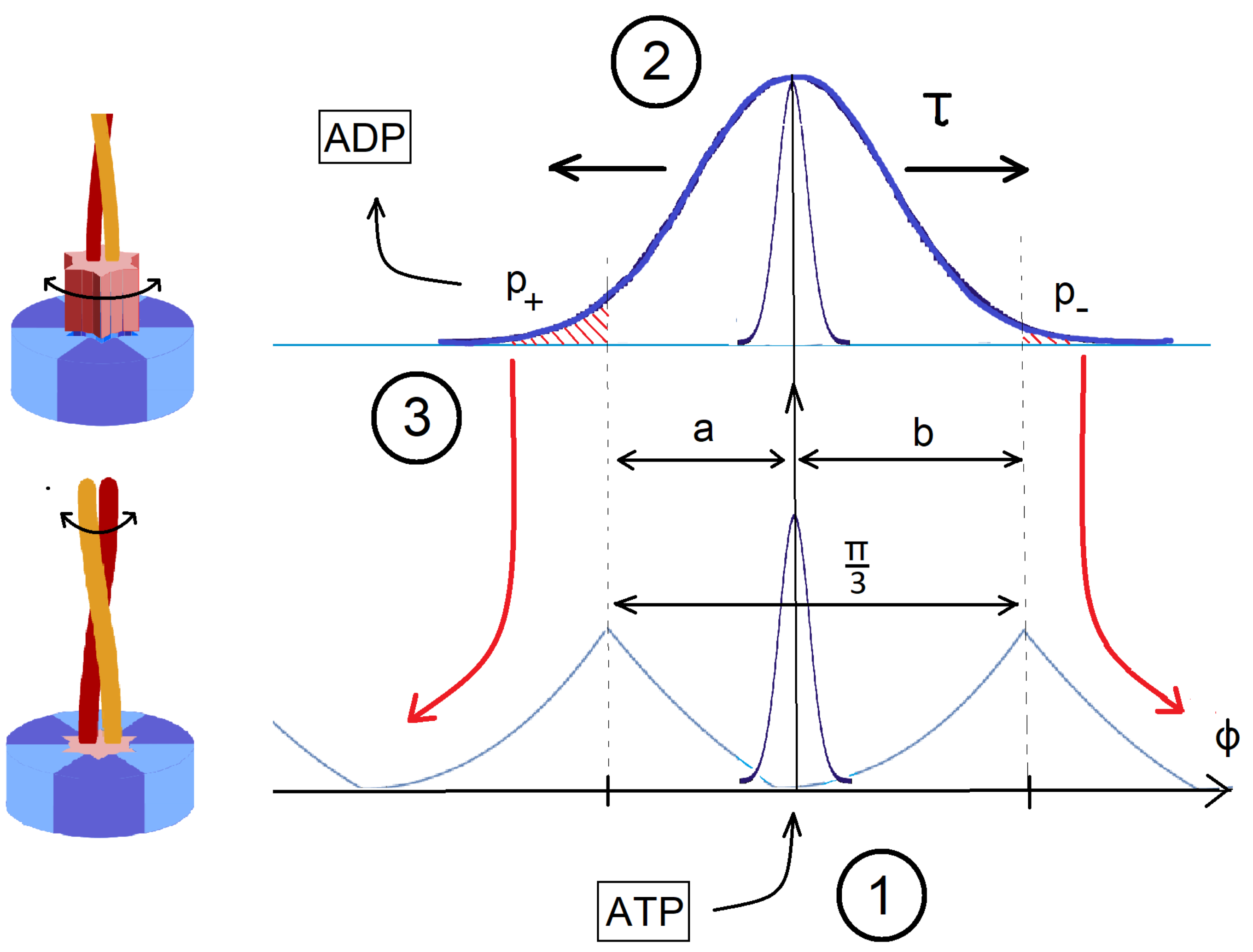}
	\caption{Motor action. (1) In the bound state, FliJ is confined to the minimum of the underlying potential seen in Fig. \ref{egraph}. When ATP binds to the FliI${}_6$ complex, the motor is excited to the second state, releasing the locked base of FliJ. (2) The coiled coil freely diffuses for a characteristic time $\tau$ that remains to be determined. (3) After that time, ADP leaves the stator and the motor collapses to the original bound state, locking the base of FliJ in the nearest available cavity. The asymmetry in the torsional potential ensures that $a<b$, and so the probability of making a step to the left $p_{+}$ is greater than the probability of making a step to the right $p_{-}$. Thus, directed motion is achieved.}
	\label{motoraction}
\end{figure}

When FliJ is bound to one of these sites, it is incapable of free rotation about its axis and is only subject to torsional fluctuations in the potential described in section II (fig. \ref{egraph}). Due to the six-fold symmetry of the stator, the potential must have a period of $\pi/3=60^\circ$. When a molecule of ATP is hydrolysed in one of the FliI subunits, the FliJ shaft is released from its confinement, and is temporarily free to rotationally diffuse (FliJ shaft freely rotating). We therefore have a two-level system described by two potentials with transitions facilitated by ATP hydrolysis (fig. \ref{motoraction}).

\section{The motor in action}
The motion of the motor is governed by two processes: the rotational diffusion of the FliJ shaft in the activated state for some characteristic lifetime  $\tau$ of the ADP-bound state, and the rate of ATP hydrolysis. that is, the excitation rate. The sequence of events that constitute a single step is described in fig. \ref{motoraction}. To calculate the average angular velocity, we can simply write:
\begin{equation}
\langle\omega\rangle=R_{ATP}\langle\Delta\phi\rangle
\end{equation}
 where $R_{ATP}$ is the rate of stepping that corresponds to the rate of ATP hydrolysis, and $\langle\Delta\phi\rangle$ is the average step size. Note that $\langle\Delta\phi\rangle\neq\pi/3$ since backward steps have a finite non-zero probability  in this model. To calculate $R_{ATP}$ we use the standard Michaelis-Menten model, in which the motor serves as a catalyst for the ATP hydrolysis:
\begin{equation}
ATP + M \rightleftharpoons [ATP\cdot M] \rightarrow ADP + M
\end{equation}
Assuming that the motor M is unchanged in the reaction and that the concentration of the ATP-motor complex $[ATP\cdot M]$ is in a steady state, one can derive the overall rate of reaction:
 \begin{equation}
R_{ATP}=\frac{V_{max}[ATP]}{K_{M}+[ATP]}
\end{equation}
where $[ATP]$ is the concentration of ATP in molars. The parameters $V_{max}$ and $K_{M}$ govern the saturation rate and half-point, respectively, and can be determined experimentally with standard hydrolysis assays. Data from Claret et al.\cite{claret2003oligomerization} give $V_{max}=233\; (mM \; \mathrm{of \, ATP})\: s^{-1}$ per one hexamer, and $K_{M}=0.65\; mM$. Here $V_{max}$ was obtained by dividing the experimentally measured $V_{max}^\mathrm{sample}$ of the whole sample by the number of FliI$_{6}$ hexamers in the assay solution. The inclusion of Michaelis-Menten kinetics is an improvement to the model proposed by Kulish et al. \cite{kulish2016f} where a simple expression $\Delta t_{step}=1/k_{on}[ATP]+\tau$ was used for the time per step, which saturates at $\tau$ when [ATP] is high. That expression is wrong because, unlike Michaelis-Menten, it omits the fact that ATP can dissociate from a FliI unit before the hydrolysis proceeds, and it also fails to include any other processes that need to occur in the motor before it is ready to take on another ATP molecule. That is, it fails to account for the `dwell time' \cite{hafner2016run} which will, in the end, be a major contributor to the time per step of rotation. 

\begin{figure}
	\centering
	\includegraphics[width=0.45\textwidth]{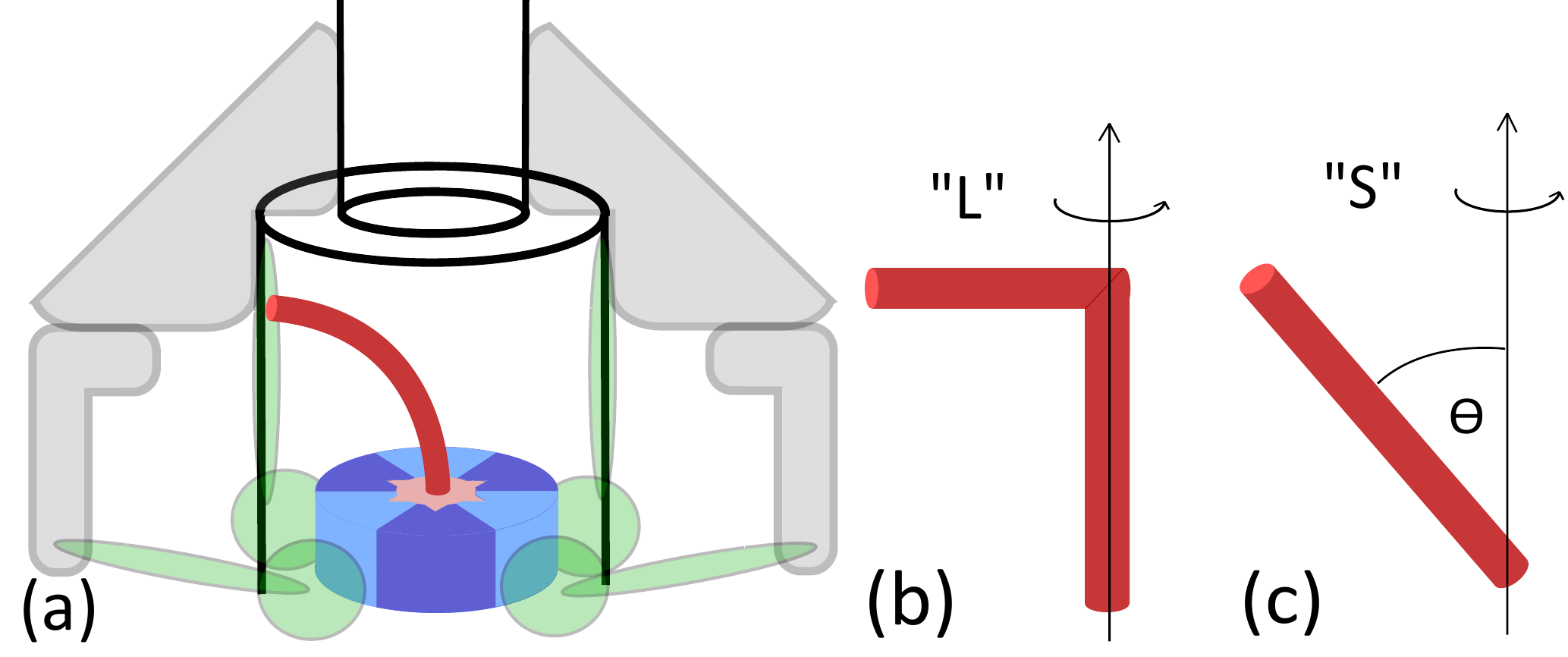}
	\caption{FliJ bending in the export apparatus. (a) In living cells, the FliJ shaft is in contact with the side of export gate, giving it a slight bend.  (b-c) Two limiting geometries considered when calculating the frictional coefficient.}
	\label{bent}
\end{figure}

The average step length $\langle\Delta\phi\rangle$ can be calculated using the 1D rotational diffusion model, and the knowledge of parameters $a\approx0.273\:rad$ and $b\approx0.774\:rad$ (the asymmetry parameters that are easily obtainable from the shape of the underlying potential, see fig. \ref{egraph}). It follows that:
\begin{eqnarray}
\langle\Delta\phi\rangle &=& \sum_{i}\langle\Delta\phi_{i}\rangle p_{i} = \frac{\pi}{3}(p_{+}-p_{-})\nonumber \\
&=&  \frac{\pi}{6}\left[\mathrm{erf}\left(\frac{b}{\sqrt{4D\tau}}\right)-\mathrm{erf}\left(\frac{a}{\sqrt{4D\tau}}\right)\right] 
\end{eqnarray}
Clearly, for $a=b$ the average velocity is zero, as must be the case in the absence of asymmetry. Two more parameters remain to be addressed, one of which is the rotational diffusion coefficient $D$ of the unrestricted FliJ filament, given by $D=k_BT/\gamma$ with $\gamma$ the friction coefficient. Structural data \cite{abrusci2013architecture} reveals that in living cells the FliJ shaft touches the side of the export channel, giving it a slight bend (fig. \ref{bent}a). Here we simplify the problem and look at the two limiting geometries: an L-shaped bar and a tilted straight bar (fig. \ref{bent}b-c). 
Rotating cylinder sets up a tangential flow around it, which has a $\propto 1/r$ velocity decay. The frictional coefficient is the proportionality constant between applied torque and rotational velocity. In a medium of viscosity $\eta$, for a cylinder of radius $R$ rotating about its axis, this is given by $\gamma=2\pi R^{2}\eta$ per unit length. This can be used to calculate $\gamma$ for the vertical part of the L shaped rod  in fig. \ref{bent}b. For low Reynolds numbers, the force per unit length acting on a cylinder moving at constant velocity $U$ perpendicular to its axis  is given approximately by \cite{shiau2014drag, lamb1911xv}:
\begin{equation} 
\frac{4\pi\eta U}{\frac{1}{2}-\Gamma-\ln{(Re/8)}} , 
\end{equation}
where $\Gamma=0.5772$ is the Euler's constant and $Re=U\rho R/\eta$ is Reynolds number. By integrating the force along the length of the cylinder we can obtain the torque and consequently the rotational frictional coefficient. Using parameters of the FliJ shaft and typical intracellular conditions ($\rho\approx\ 1100\;kg\:m^{-3}$; $\eta\approx 10^{-3}\;Pa\:s$ and $T\approx300\;K$) we finally arrive at the two limiting diffusion coefficients $D_{L}=1.38\times 10^7\; \mathrm{rad}^{2}\:s^{-1}$ and $D_{S}=1.53\times 10^7\;\mathrm{rad}^{2}\:s^{-1}$ (for $\theta=45^\circ$). These values are very close and henceforth $D_{L}$ was used in all calculations. 

\begin{figure}
	\centering
	\includegraphics[width=0.35\textwidth]{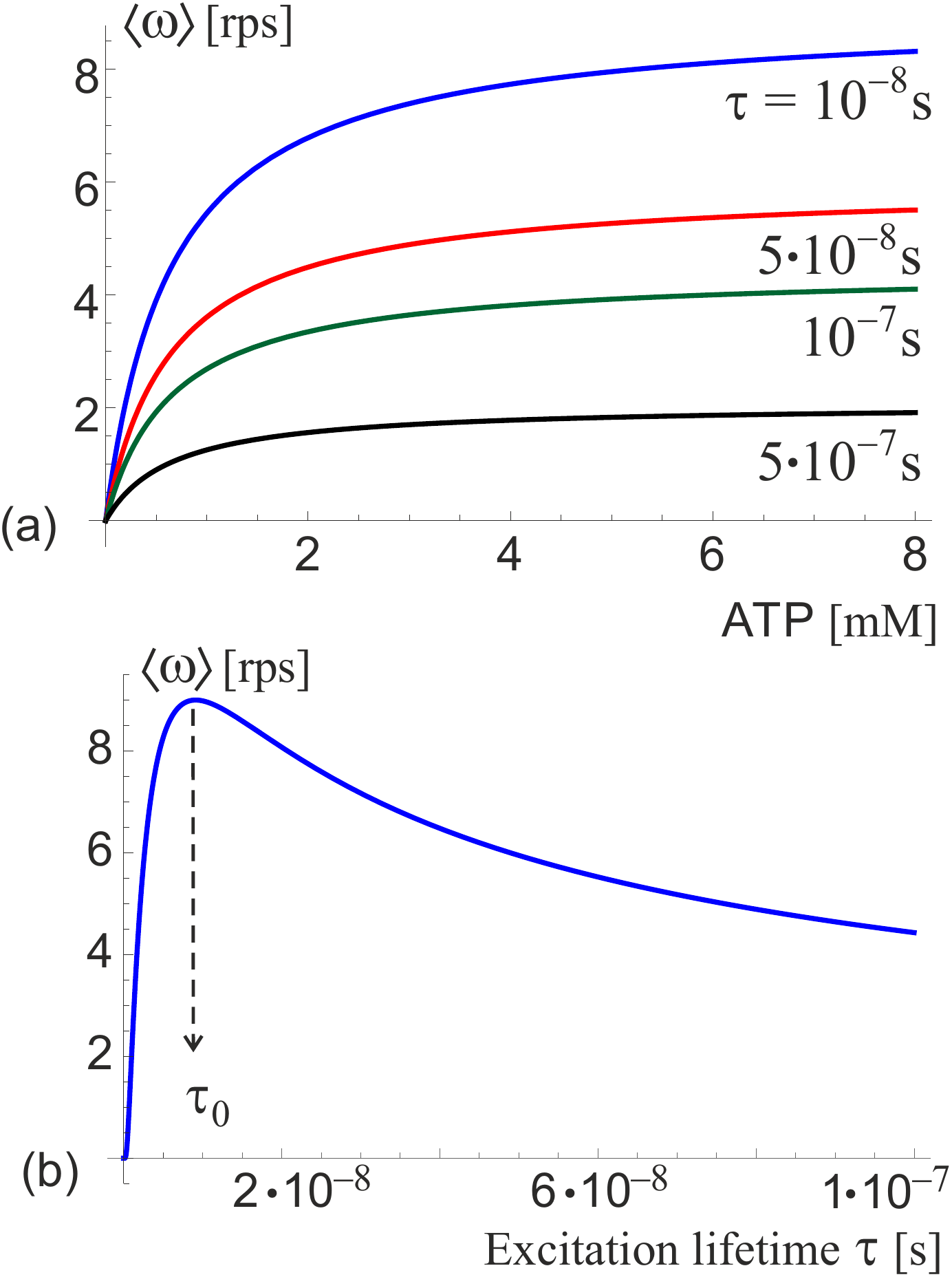}
	\caption{Average angular velocity of the free ATPase motor. (a) Angular velocity as a function of ATP concentration, plotted for several values of the excitation lifetime $\tau$ labelled on the plot. Enzyme kinetics ensures that the velocities saturate at high concentrations. (b) Angular velocity as a function of $\tau$ in the ATP-saturated regime. A maximum is attained at $\tau_{0}\approx 8\times 10^{-9}s$. which corresponds to $\omega_\mathrm{max}\approx 9.0\:rps$.}
	\label{graphs}
\end{figure}

We can now write the full expression for the average angular velocity:
\begin{equation}
\omega=\frac{V_{max}[ATP]}{K_{M}+[ATP]}\frac{\pi}{6}\left(\mathrm{erf}\left(\frac{b}{\sqrt{4D\tau}}\right)-\mathrm{erf}\left(\frac{a}{\sqrt{4D\tau}}\right)\right)
\label{wfree}
\end{equation}
The dependence of $\omega$ on ATP concentration for several values of $\tau$ is plotted in fig. \ref{graphs}a. Notice the saturation at high ATP concentration due to Michaelis-Menten kinetics. $\tau$ is a free parameter and has a significant effect on the average (figure \ref{graphs}b). In order to infer key features of the motor a specific value of $\tau$ needs to be chosen. In this study, the velocity-optimal value $\tau_{0}\approx 8\times 10^{-9}s$ was chosen, which corresponds to the maximum average angular velocity $\omega_\mathrm{max}\approx9.0\:rps$. Determination of factors affecting the excitation lifetime is beyond the scope of this project but is necessary to gain a full understanding of the system and should, therefore, be subject of further studies.

\section{ ATP$\mathrm{\bf{ase}}$ under external torque}
`Viscous resistance' for the FliJ rotation has already been accounted for using the diffusion coefficient $D$. In the previous case of free motor spinning, i.e. in the absence of an external torque, the power output of the motor is zero by definition. All energy released by ATP hydrolysis is eventually converted into movement of the surrounding molecules and there is no 'useful' work done. When an external torque $G$ is applied to the motor, both the ground state and the excited state potentials are adjusted by $\Delta U=-G\phi$. This has two effects: it changes the asymmetry parameters $a$ and $b$ of the underlying potential of the ground state, and it affects the diffusion in the excited state. It is useful to introduce the concept of a stall torque $G_\mathrm{stall}$, which is the counter-torque when the average angular velocity $\omega=0$. When a constant torque is applied to a freely diffusing system with initial condition $P(\phi, t=0)=\delta (\phi)$, the probability distribution has the following form including the drift term:
\begin{equation}
P(\phi, \tau)=\frac{1}{\sqrt{4\pi D\tau}}\exp{\left(-\frac{(\phi-G\tau/\gamma)^{2}}{4D\tau}\right)}
\end{equation}
so the origin of diffusion shifts in time by $\delta\phi=G\tau/\gamma$. This in return effectively changes the asymmetry parameters $a\rightarrow a+\delta\phi$ and $b\rightarrow b-\delta\phi$. According to Eq. \eqref{wfree} when $a+\delta\phi=b-\delta\phi$, we can solve the condition $\omega=0$ which yields the stall torque:
\begin{equation}
G_\mathrm{stall}=\frac{k_BT}{D\tau}\frac{b'-a'}{2}  \label{Gs1}
\end{equation}
The primes denote the fact that the asymmetry parameters $a=a(G)$ and $b=b(G)$ are also functions of the applied torque, because the torque changes the shape of the underlying torsional potential energy $E(\phi)$ depicted in fig. \ref{egraph}, adjusted by the external torque:  $E_{n}=E(\phi)-G\phi$. Periodicity requires that $E_{n}(b')=E_{n}(-a')$ and $a'+b'=\pi/3$ as before. We can expand $a'=a-\Delta$ , $b'=b+\Delta$ for small alterations $\Delta$ to obtain an expression for the shift of the asymmetry $\Delta \approx \frac{\pi}{3}G/[E'(b)-E'(-a)]=\frac{\pi}{3}G/\Delta E'$. By plugging this back into Eq. \eqref{Gs1}, we finally obtain the  condition for stall torque:
\begin{equation}
G_\mathrm{stall} \approx {k_BT}\, \frac{b-a}{2}\frac{\Delta E'}{D\tau\Delta E' - \frac{\pi}{3}k_BT} .
\end{equation}

\begin{figure}
	\centering
	\includegraphics[width=0.35\textwidth]{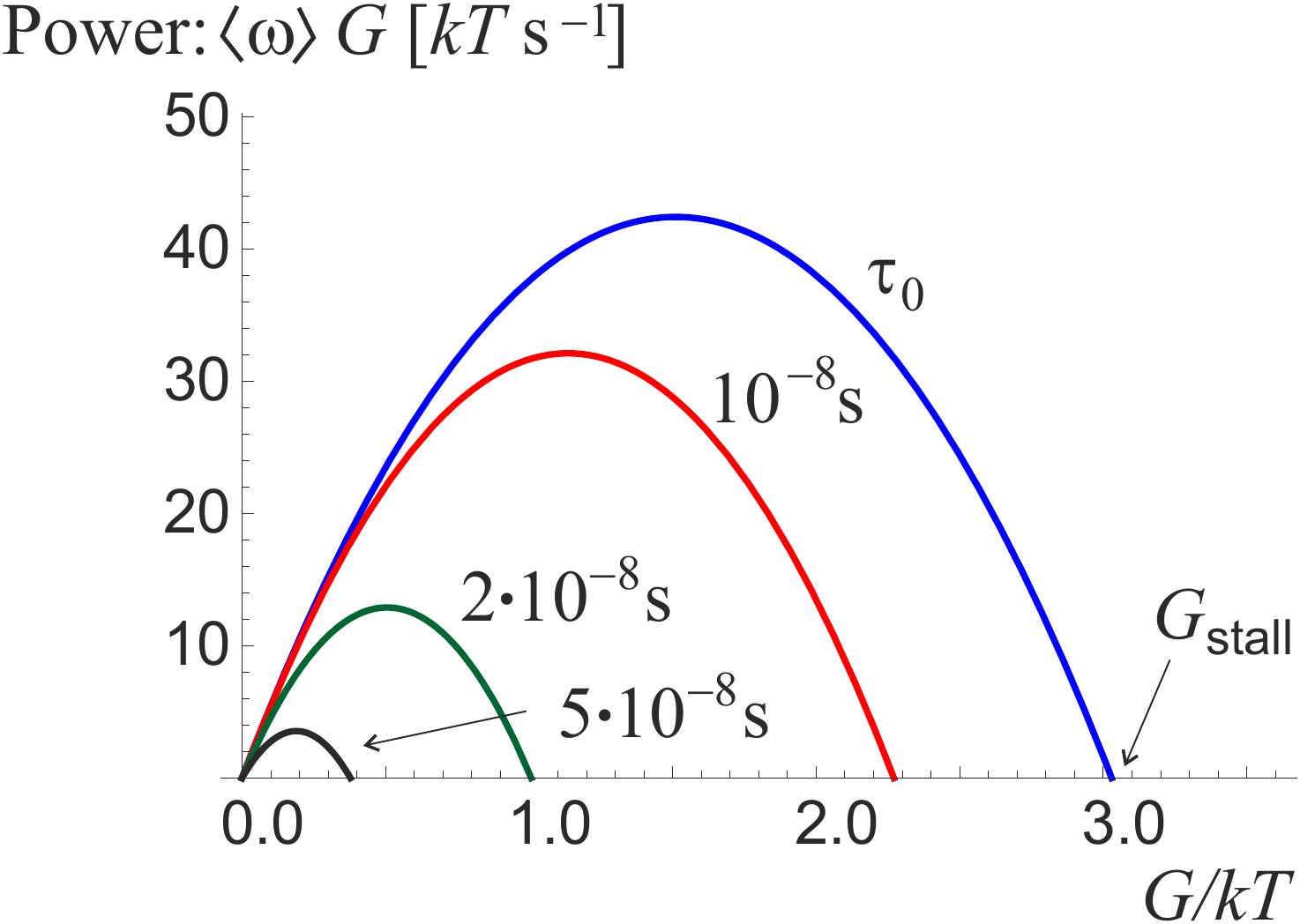}
	\caption{The power output of the motor as a function of the external torque for several values of $\tau$. The power goes to zero at $G_{stall}$ when $\omega=0$. For each $\tau$ there is a maximum power attained. Taking $\tau_{0}$ as the reference, the values are $G_\mathrm{stall}\approx 3 \, k_BT\: \mathrm{rad}^{-1}$; $P_\mathrm{max}\approx 42 \, k_BT\: s^{-1}$}
	\label{power}
\end{figure}

The overall power output of the motor is given by $P_\mathrm{out}=G \langle \omega(\tau, G) \rangle$. The output power is never negative because the torque is provided by the resistance of the surroundings, which can only stop the motor but not force it to rotate in the opposite direction. This is different from the case of $F_1 F_0$ ATPase where the $F_0$ motor can provide enough counter-torque to reverse the sense of rotation and switch from ATP hydrolysis to synthesis \cite{kulish2016f}.  Using the optimal-velocity value $\tau_{0}$ discussed in the case of free-spinning motor (fig. \ref{graphs}b), we can infer the magnitude of stall torque, as well as the maximum power output $P_\mathrm{max}$, see fig. \ref{power}. 
The parameters of the motor are $G_\mathrm{stall}\approx 3 \, k_BT\: \mathrm{rad}^{-1}$ and $P_\mathrm{max}\approx 42 \, k_BT\: s^{-1}$ at $G_\mathrm{max}\approx 1.5 \, k_BT\:\mathrm{rad}^{-1}$. By taking the ratio $\eta_\mathrm{max}=P_\mathrm{max}/P_\mathrm{in}$, where the input power $P_\mathrm{in}=R_{ATP}\Delta G_{ATP}$ is the rate of energy released by ATP hydrolysis, with $\Delta G_{ATP}\approx21 \, k_BT$ and $[ATP]\approx 1\, mM$ as some typical  cellular conditions  \cite{kinosita2000,kulish2016f}, we arrive at the maximum efficiency $\eta_\mathrm{max}\approx1.4\%$. Despite being surprisingly low, the efficiency is physically plausible and corresponds to a temperature difference of $\Delta T=4.2K$ between a hot and a cold reservoir of a Carnot engine -- a temperature difference practically attainable within living organisms.

\section{Assisted unfolding}

The exact role of the FliI${}_6$-FliJ complex at the base of the bacterial flagellar export apparatus remains unclear despite the detailed structural data  available. The diameter of the export channel, $d \approx 2$ nm implies that the flagellin subunits that eventually assemble into a flagellum outside of the cell need to be unfolded first in order to pass through such a narrow channel. Minamino  \cite{minamino2014protein} states that this is achieved with the use of proton motive force (pmf), and ATP activity of the FliI${}_6$-FliJ complex. It is not clear to us where the pmf could contribute to this process. On the other hand, several studies suggest \cite{bai2014assembly, minamino2011energy} complex process of ``loading'' of the subunit into the channel by binding to chaperone proteins that assist in anchoring and unfolding of the subunit. It has also been shown that even in the absence of ATPase activity certain level of protein export was achieved \cite{minamino2014bacterial}, suggesting that the motor is not critical for the export and merely \textit{assists} in the process. We, therefore, propose that the purpose of the FliI${}_6$-FliJ motor complex  is in the \textit{assisted unfolding} of the substrate. 

\begin{figure}
	\centering
	\includegraphics[width=0.30\textwidth]{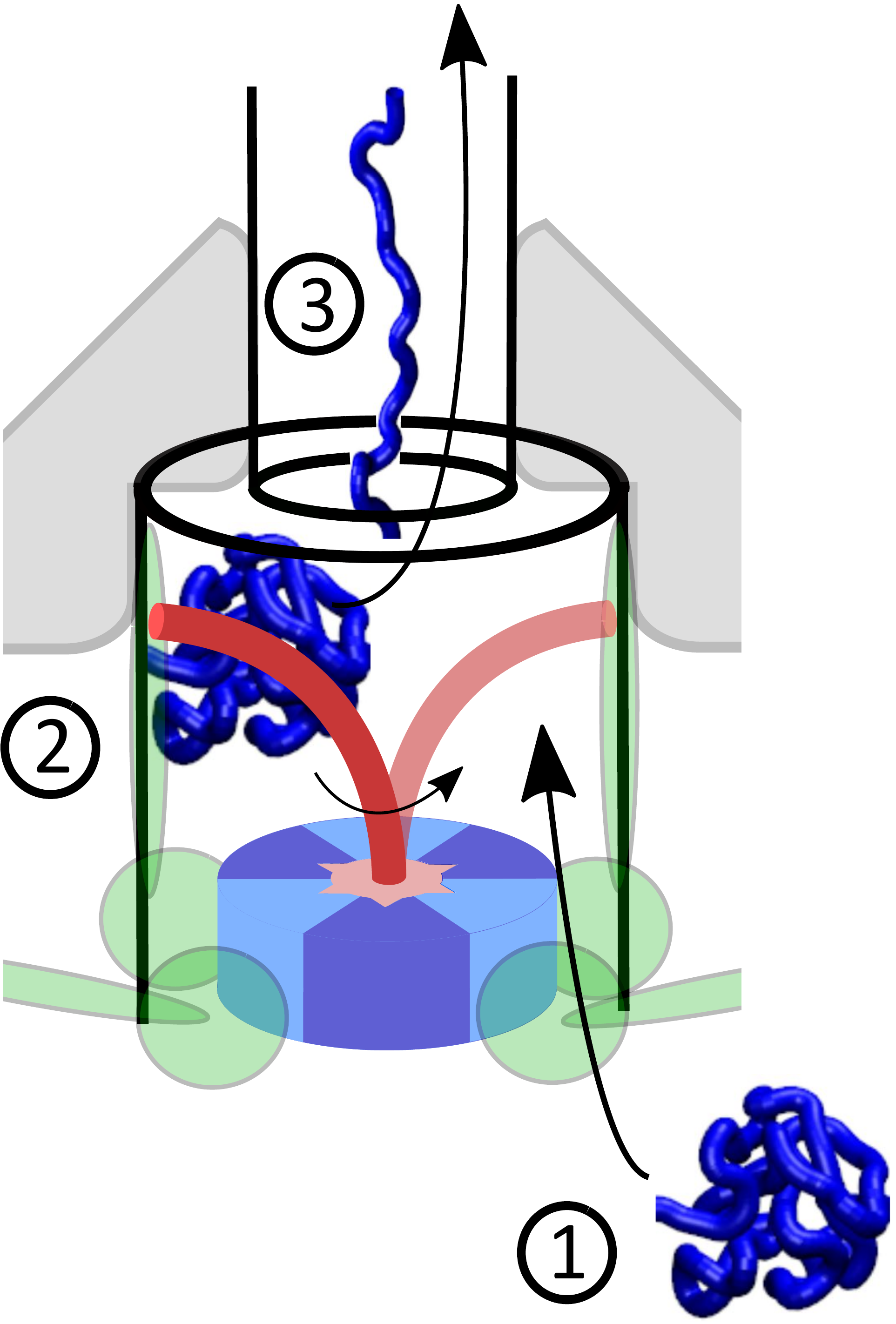}
	\caption{Assisted unfolding. (1) Folded protein arrives at the base of the export channel where the FliI$_{6}$-FliJ motor is situated and enters the gate area. (2) Rotating FliJ unit hits the substrate, providing enough energy to help unfold the protein. (3) Unfolded protein is then funnelled into the channel and with the use of proton motive force transported to the exterior of the cell.}
	\label{unfolding}
\end{figure}

This is achieved via direct mechanical interaction of the subunits  with the rotating FliJ filament (fig. \ref{unfolding} illustrating the point). The subunit, which is temporarily bound to the cavity, resists the filament movement, the associated counter-torque slowing down its natural rotation in the motor. The motor exerts an equal and opposite force on the substrate, helping it overcome the energy barrier and unfold (a number of studies give details of protein unfolding under force \cite{gaub1999,fernandez2007,bell2016}). 

Two possible modes of action are distinguished. In the \textit{stall mode} the rotation of the motor is fully suppressed by the substrate which is able to apply $G_\mathrm{stall}$ to the motor. This counter-torque builds up tension in the subunit protein, until it unfolds with the help of other mechanisms such as chaperones \cite{stryer_1999, alberts_2015}, or just stochastically  \cite{fernandez2007,bell2016}. Mean first-passage time can be used to calculate the average time for such stochastic unfolding.   In contrast, the \textit{grinding mode} could occur instead, in which the motor continuously rotates (grinds through, or past the bound substrate) and supplies the additional energy to the unfolding. Using simple reaction kinetics, the unfolding rate $r\propto \exp{(-E_\mathrm{unfold}/k_BT)}$. We can approximate the additional energy supplied by the motor as $\Delta E\approx\delta\phi G_\mathrm{max}\approx\pi G_\mathrm{max}/3$ per step. So the overall unfolding rate is increased by a factor $\exp{(-\Delta E/k_BT)}\approx 4.8$. This result is in very good agreement with experiments performed by Minamino et al.\cite{minamino2008distinct} and Paul et al. \cite{paul2008energy}, where the FliI ATPase was removed or damaged by a mutation. These papers report a three- and four-fold reduction in flagellar secretion, respectively, upon removal of the FliI${}_6$-FliJ complex. Our slight overestimate of the rate reduction is mainly due to the assumption that all the work performed by the motor is utilised in unfolding the substrate (when there would naturally be losses). The agreement with experimental values shows that the proposed ``Brownian ratchet'' model is a viable alternative way of describing rotary molecular motors.

\section{Discussion}
In this study, we attempted to model a rotational molecular motor exploiting the mechanism already known to be working for linear motors \cite{julicher1997modeling}. We took the example of the FliI${}_6$-FliJ ATPase, whose function in the bacterial flagellar export apparatus is still unknown, to predict its rotation speed and power output, and ascertain its functionality. 

The two-state model  \cite{julicher1997modeling, kulish2016f} was used to describe the average outcome of the stochastic rotary motion. In the ground state, the underlying potential derives from the characteristically asymmetric torsional elastic properties of the FliJ coiled coil. This was obtained by assuming two elastic circular rods (representing $\alpha$-helices) intertwined together by sharing the hydrophobically-bonded seam line. Such a model is a large simplification: in reality, the helices are not of equal length, and one cannot assume uniform bending and radius of the coiled coil along the whole filament because of side chains that protrude out of the helices. Molecular dynamics simulation might give a more accurate estimate of the torsional elastic properties as well as of the hydrophobic interaction and would, therefore, be needed for further development of this model. 

In the excited state, the potential confining the coiled-coil filament is flat, and instead of twisting the FliJ filament undergoes free rotational diffusion. Here the diffusion coefficient was approximated as that of a uniform cylinder moving through a fluid at low Reynolds number. This approximation is necessary to obtain any numerical estimates and could be corrected in the future by performing measurements on flow in an optical trap. However, we believe our estimates of the friction and rotational diffusion constant are reasonably accurate, as they match many experimental studies of coiled-coil rotation in a cavity of protein complex \cite{yasuda1998f1, kulish2016f}. 

Transitions between states are induced by the ATP hydrolysis, and the asymmetry of the ground state torsional potential gives rise to directed motion. The excitation lifetime $\tau$ is a free parameter in this model, and it significantly affects key features of the motor. A value $\tau_{0} \approx 8\: ps$ was chosen that maximizes the average angular velocity of free rotation, in order to obtain numerical estimates of the stall torque, power output, etc. but a more careful choice of $\tau$ should be made in further studies based on external factors and physical limitations of the system.   

Despite the numerous approximations that were made, qualitatively good results were obtained for the average angular velocity $\omega_{max}\approx9.0\:rps$, stall torque $G_{stall}\approx 3.0\:k_BT\:rad^{-1}$ and maximum power output $P_{max}\approx 42.0\:k_BT\: s^{-1}$ that match many experimental data, and suggest that the FliI${}_6$-FliJ motor might assist in mechanical unfolding of the proteins that are subsequently exported and assembled into a flagellum. This proves the viability of our approach to modelling the motion of such molecular motor. Naturally, a possible next step could be to account for any internal degrees of freedom and transition from an overly simplified two-state model to an N-state model that captures the complex nature of soft matter systems more accurately.

\section*{References}
%

\end{document}